\documentclass[10pt,conference]{IEEEtran}
\IEEEoverridecommandlockouts
\usepackage{cite}
\usepackage{multirow}
\usepackage{graphicx}
\usepackage{rotating}
\usepackage[export]{adjustbox}
\usepackage{amsmath,amssymb,amsfonts}
\usepackage{algorithmic}
\usepackage{graphicx}
\usepackage{afterpage}
\usepackage{textcomp}
\usepackage{pdfpages}
\usepackage{caption}
\usepackage{subcaption}
\usepackage{xcolor}
\usepackage{placeins}
\usepackage{multicol}
\usepackage{url}
\usepackage{lipsum}  
\usepackage{verbatim}
\usepackage[inline]{enumitem}
\usepackage{setspace}
\usepackage{float}
\usepackage{tabularx}
\usepackage{adjustbox}
\usepackage{array}
\usepackage{enumitem}   
\usepackage{array}
\usepackage{multirow}

\newcolumntype{P}[1]{>{\centering\arraybackslash}p{#1}}
\renewcommand\thesubsection{\Alph{subsection}}

\begin{document}

\title{Improving HTTP/3 Quality of Experience with Incremental EPS}

\author{\IEEEauthorblockN{\textsuperscript{} Abhinav Gupta and Radim Bartos}
\IEEEauthorblockA{\textit{Department of Computer Science} \\
\textit{University of New Hampshire}\\
Durham, NH 03824, USA \\
{\{ag1226,rbartos\}}@cs.unh.edu}}

\maketitle

\begin{abstract}

With the introduction of QUIC, a modern transport-layer network protocol, HTTP/3 leverages its benefits to enhance web content delivery. This paper proposes a mechanism based on the recently standardized Extensible Prioritization Scheme (EPS) for weighted incremental web content delivery. The mechanism augments the sequential scheduling to provide incremental and weighted incremental resource delivery. An existing HTTP/3 implementation was extended with the proposed mechanism and tested with the content of eight popular websites. The results of our experimental analysis show that weighted incremental prioritization improves Quality of Experience (QoE) as measured by Lighthouse, a standard QoE test tool. While overall improvements were generally achieved, we also observed a few cases where the performance degraded slightly, highlighting that the QoE is sensitive to factors such as web page structure.

\end{abstract}

\begin{IEEEkeywords}
Extensible Prioritization Scheme, HTTP/3, QUIC, QoE, Lighthouse, Protocol Performance
\end{IEEEkeywords}

\section{Introduction}

The call for more robust network protocols intensifies as the digital world expands to improve internet services. At the forefront of meeting this demand is the continual evolution of HTTP, which has become a cornerstone in enhancing the Quality of Experience (QoE) for users around the globe.

The most recent version, HTTP/3, offers multiple improvements over previous versions, including reduced connection establishment times and a decrease in head-of-line blocking, among other enhancements. This leap in protocol efficiency largely stems from HTTP/3 utilizing QUIC, leveraging its stream multiplexing features to accelerate web communication speeds significantly. In tandem with these developments, the Extensible Prioritization Scheme (EPS)~\cite{rfc9218} emerged, marking a paradigm shift in resource prioritization strategies. EPS replaces the complex dependency tree with a system comprising of eight stream urgency levels, and the incremental parameter that indicates whether a resource should be delivered sequentially (RFC 9218 \cite{rfc9218} refers to sequential delivery as non-incremental) or incrementally together with other such resources.



Prioritization dictates the sequence and method by which web content is sent. By leveraging QUIC to transfer data across multiple streams, prioritization strategies are finely tuned to accelerate the delivery of essential website components. Many HTTP/3 implementations~\cite{same_standards} adopt sequential or round-robin delivery, which can lead to head-of-line blocking. These common strategies do not always align with the demands of modern web traffic and user expectations~\cite{not_good_roundrobin,sander2022analyzing}.

To assess the QoE of a webpage delivery, various web performance metrics~\cite{core_web_vitals} serve as indicators,  capturing the user experience from the beginning of page loading to their interaction with the webpage. 

In this paper, we assess the QoE by using metrics provided by Lighthouse~\cite{lighthouse}, an open-source, automated tool developed by Google for measuring the performance of web pages. Lighthouse has been used in previous web performance studies~\cite{saif2020early,gupta}. First Contentful Paint (FCP) measures the time it takes for the first visual content of a page to load. Following this, the Largest Contentful Paint (LCP) times the rendering of the most significant visual element, contingent upon the FCP. Speed Index (SI) encapsulates the rate of visual content delivery, drawing on the FCP and LCP for a comprehensive measure. Time to Interactive (TTI) measures the point at which a page becomes fully interactive, pending the occurrence of FCP, while Total Blocking Time (TBT) quantifies the duration of main thread inactivity post FCP.
These first five metrics are time-based, meaning lower values indicate better performance. In contrast, Cumulative Layout Shift (CLS) is a unitless measure. CLS assesses the frequency and magnitude of visual instability, operating independently from the other metrics.

The web page structure influences various QoE metrics, as described above. For instance, pages with simpler structures and content load faster, positively affecting FCP and LCP scores. Conversely, complex pages with numerous interacting elements and scripts may increase TBT and delay TTI, compromising user satisfaction. Additionally, design that promotes stable layout can minimize CLS.


\section{Related Work}
\label{sec:related_work}
The evolution of web performance research has been significantly shaped by ongoing HTTP/3 and QUIC performance evaluations~\cite{dissecting_QUIC,QUICDesignInternetScale,WebQUICFaster,LonglookatQUIC}, which have shed light on diverse protocol implementations and their influence on QoE.


Foundational studies focusing on web resource prioritization have significantly enriched our understanding in this domain~\cite{rmarx_resource,sander2022analyzing,h2_wijnants,ResourceMultiplexing,meenan_http2_prioritization}. The advent of standards like EPS (RFC 9218)~\cite{rfc9218}, alongside HTTP/3~\cite{rfc9114} and QUIC~\cite{rfc9000} extends the groundwork of web resource prioritization and indicates opportunities for future research in this field.

Wijnants et al.~\cite{h2_wijnants} find that HTTP/2 prioritization practices vary widely among major browsers, affecting webpage loading efficiency. Their studies show complex prioritization methods generally outperform simpler ones, with the latter potentially slowing down median visual load times by over 25\%. Furthermore, they suggest that server-side re-prioritization is a complex task heavily influenced by browser-specific behaviors.

Improving QoE extends beyond HTTP prioritization; a substantial amount of research is being directed toward techniques that offer alternative solutions. These techniques include server push~\cite{http2_server_push,how_spdy_is_spdy}, resource hints like preload~\cite{patrick_http2}, and custom JavaScript based scheduler~\cite{netravali}, optimal browser heuristics~\cite{page_load_performance} all aimed at optimizing how resources are managed and delivered.

Marx et al.~\cite{ResourceMultiplexing} investigated the efficacy of various multiplexing behaviors and prioritization schemes in HTTP/3, revealing substantial performance disparities that can lead to up to fivefold differences in page load times under certain conditions. Their investigation highlights the context-sensitive nature of these improvements, heavily influenced by web page structure and network environments. They advocate for simplified prioritization frameworks and highlight the adoption of the EPS.


Furthering this exploration, the subsequent study by Sander et al.~\cite{sander2022analyzing} provides deeper insights into how these prioritization strategies manifest in web performance. Sanders et al. investigation reveals that parallel scheduling can mitigate head-of-line blocking under high random loss conditions. Yet, this strategy does not lead to consistent performance improvements across various web architectures, as website architecture substantially influences performance outcomes. 
Specifically, they observe that a combined approach of resource prioritization and parallelism is more effective than a pure round-robin for moderate loss scenarios, even with the implementation of a simplified strategy like the EPS.

The goal of this project is to study the impact of incorporating urgency levels within an incremental resource delivery mechanism. 


\section{Proposed mechanism}
\label{sec:proposed_mechanism}

Delivering web content in sync with the requirements of a browser's rendering engine ensures that every web page component is delivered when needed and avoids delays in displaying the content to the user. We propose to achieve this just-in-time delivery by integrating the proposed mechanism within an HTTP/3 protocol implementation. 

To enhance web content delivery, this paper introduces the weighted incremental scheduling mechanism. This mechanism processes the urgency levels associated with various web resources, as outlined by the EPS~\cite{rfc9218}, and calculates bandwidth share for each resource. 
This allocation promotes an orderly and effective loading sequence, ensuring that crucial resources are given priority and reach the browser quickly. With this method, we target improvements in crucial web performance metrics, such as FCP and LCP.



The design of our bandwidth share mechanism, specifically tailored to ensure effective and fair distribution of resources, is guided by three key objectives:

\begin{enumerate}
 \renewcommand{\labelenumi}{(\roman{enumi})} 
    \item \textbf{Prioritization of Urgency:} Requests with higher urgency levels should receive a more significant share of bandwidth.
    \item \textbf{Equality at the Same Urgency Level:} Within the same level of urgency, each request should be allocated an equal share of bandwidth, promoting fairness among requests of similar importance.
    \item \textbf{Proportional Bandwidth Allocation:} The bandwidth allocation between streams at different urgency levels should be proportional to their urgency. This means that the bandwidth distribution ratio between any two streams should correspond to the ratio of their urgency levels. 
    
\end{enumerate}

Building upon these objectives, the central focus of our mechanism's design is to achieve a balance that maintains the original bandwidth proportions dictated by resource urgency levels. A critical aspect of this balance is the prevention of starvation among requests, particularly for those with lower urgency. To effectively address these challenges, our mechanism integrates a carefully crafted formula, ensuring that the distribution of resources aligns with the outlined objectives.

The mechanism assigns bandwidth shares based on a combination of urgency for each request and the ratio of the same urgency level requests. To complement our bandwidth allocation objectives and enhance the flexibility of our mechanism, we introduce a weight factor $\alpha$. 

The motivation for introducing $\alpha$, $0 \leq \alpha \leq 1$, is to explore the spectrum between priority-driven bandwidth allocation and uniform distribution. It is a tuning parameter to balance between incremental and weighted incremental resource distribution. At one end of the spectrum ($\alpha = 0$), it yields an equal bandwidth distribution regardless of urgency, ensuring complete incremental resource delivery, which corresponds to the round-robin delivery, a benchmark strategy commonly employed in HTTP/3 implementations\cite{same_standards}. 

At the other end of the spectrum ($\alpha = 1$), the formula prioritizes requests strictly based on urgency. This range of $\alpha$ allows for adjustments according to the specific needs of the web page. In scenarios where urgency levels are equal, the formula guarantees that bandwidth is shared equally, ensuring fairness. The initial weight $w'_i$ of a resource $i$ is calculated based on its urgency level and the proportion of the same urgency requests:
\begin{equation}\label{eq:equation1}
w'_i =  \alpha \frac{1}{u_i + r_u} + (1-\alpha) \frac{1}{n} 
\end{equation}

\noindent
where \( u_i \) is the EPS urgency level~\cite{rfc9218} of request $i$, \( r_u \) is the ratio of the number of requests with urgency level \( u \) to the total number of requests, and $\alpha$ is the weight factor described above. To obtain bandwidth shares $w_i$ across all $n$ resources, we normalize initial weights:


\begin{equation}
w_i = \frac{w'_i}{\sum_{j=1}^{n} w'_j}
\end{equation}

\section{Experimental Setup}
\label{sec:experimental_setup}


In this study, we use \emph{aioquic}~\cite{aioquic}, a Python-based,  HTTP/3 and QUIC library. We augment this library with EPS~\cite{rfc9218} and our proposed weighted incremental mechanism to optimize HTTP/3 QoE performance.

\begin{figure}[t]
  \includegraphics[width=\linewidth]{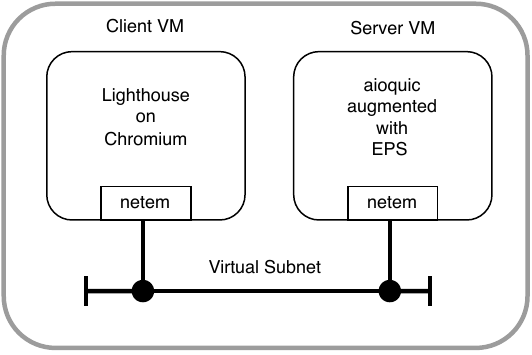}
  \caption{Experimental Setup}
  \label{Experimental_Setup}
\end{figure}

Our experimental setup consists of a Client Virtual Machine (VM) and a Server Virtual Machine (VM). Each runs Ubuntu 22.04.2 on a Linux 5.15.0-76-generic kernel, as shown in Figure~\ref{Experimental_Setup}. We orchestrate this setup using Parallels Desktop 19 for Mac Pro Edition~\cite{parallels}. The host machine was a MacBook Pro with macOS Sonoma version 14.0, equipped with 64GB of RAM and an Apple M1 CPU.

Client and Server VMs operate and communicate over a dedicated virtual subnet and are not connected to the internet during the experiments. To emulate content delivery from a Content Delivery Network (CDN), \emph{netem} is applied to the network interfaces of both Virtual Machines (VMs), introducing a loss of 0.05\% and a latency of 10 ms in each direction.

\begin{figure}[t]
  \includegraphics[width=\linewidth]{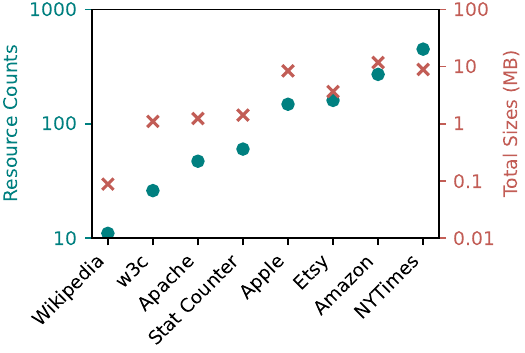}
  \caption{Number and Size of Downloaded Websites}
  \label{websites_with_resources}
\end{figure}

Server VM hosts an augmented \emph{aioquic}-based 
HTTP/3 server. On the Client VM, Lighthouse version 11.2.0, running on Chromium~\cite{Chromium} version 118.0.5993.70, is used for performance evaluation. To accurately gauge the performance metrics, we selected a set of eight websites with diverse number of resources and resource sizes, as shown in 
Figure~\ref{websites_with_resources}. 

Each website was downloaded in October 2023, and the content was stored locally. To ensure a controlled test environment, external trackers and references were removed from a few of these websites to prevent downloading additional content and maintain the isolation of our testing setup.

\begin{table}[tbh]
\centering
\caption{Mapping from Chromium to EPS}
\label{table_mapping}
\begin{tabular}{ll}
\hline
\textbf{Chromium Priority} & \textbf{EPS Urgency} \\ \hline
Very High (0) & 0  \\
High (1)      & 2 \\
Medium (2)    & 3 (EPS Default)\\
Low (3)       & 5 \\
Very Low (4)  & 7 \\ \hline
\end{tabular}
\end{table}

The experiments were run for five values of $\alpha$: 0, 0.25, 0.5, 0.75, and 1. Ten iterations of each Lighthouse run per website were conducted. During these tests, we recorded six metrics: FCP, LCP, TTI, SI, TBT, and CLS.

Chromium did not send Priority HTTP header field or PRIORITY\_UPDATE frames as observed on the \emph{aioquic}-based HTTP/3 server. Due to the absense of these priority signals, we extracted Chromium priorities from Lighthouse reports and mapped them to EPS urgency levels, as shown in Table~\ref{table_mapping}.

\begin{figure}[t]
  \includegraphics[width=\linewidth]{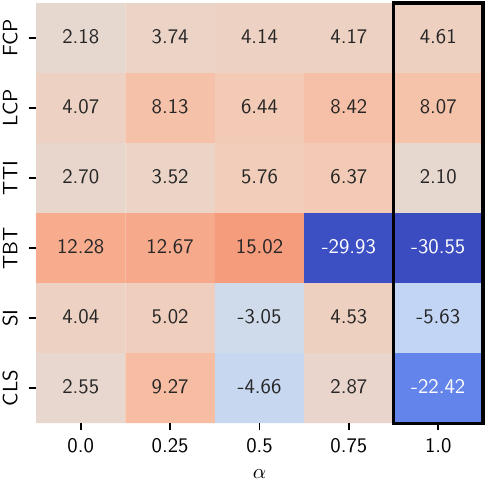}
  \caption{Percentage Metric Improvements over Sequential Delivery for Varying  $\alpha$ Values}
  \label{heatmap}
\end{figure}

\begin{figure}[t]
  \centering
  \begin{subfigure}[b]{\linewidth} 
    \includegraphics[width=\linewidth]{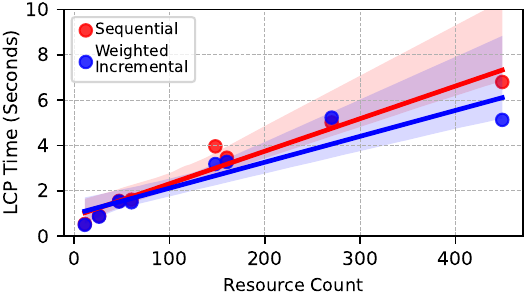}

    \caption{Sorted by Resource Count\hspace*{-5em}}
    \label{lcp_1}
  \end{subfigure}\\[15pt] 
  
  \begin{subfigure}[b]{\linewidth} 
    \includegraphics[width=\linewidth]{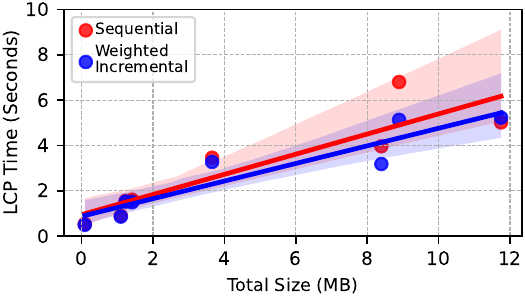}
\captionsetup{justification=justified}
    \caption{Sorted by Total Size\hspace*{-3.5em}}
    \label{lcp_2}
  \end{subfigure}\\[10pt]
  \caption{LCP Improvement Trends ($\alpha = 1$)}
  \label{regression} 
\end{figure}

\begin{figure*}[htbp!]
  \centering
  \newlength{\figheight}
  \setlength{\figheight}{.18\textheight} 
  \begin{subfigure}[b]{0.44\textwidth}
    \includegraphics[width=\textwidth, height=\figheight, keepaspectratio]{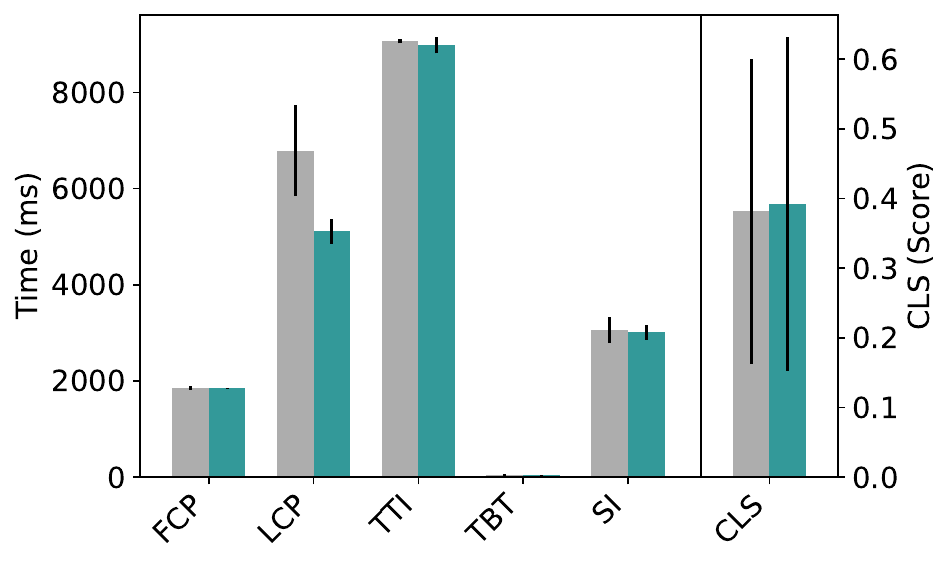}
    \caption{www.amazon.com}
    \label{amazon}
  \end{subfigure}
  \hspace{0.02\textwidth} 
  \begin{subfigure}[b]{0.44\textwidth}
    \includegraphics[width=\textwidth, height=\figheight, keepaspectratio]{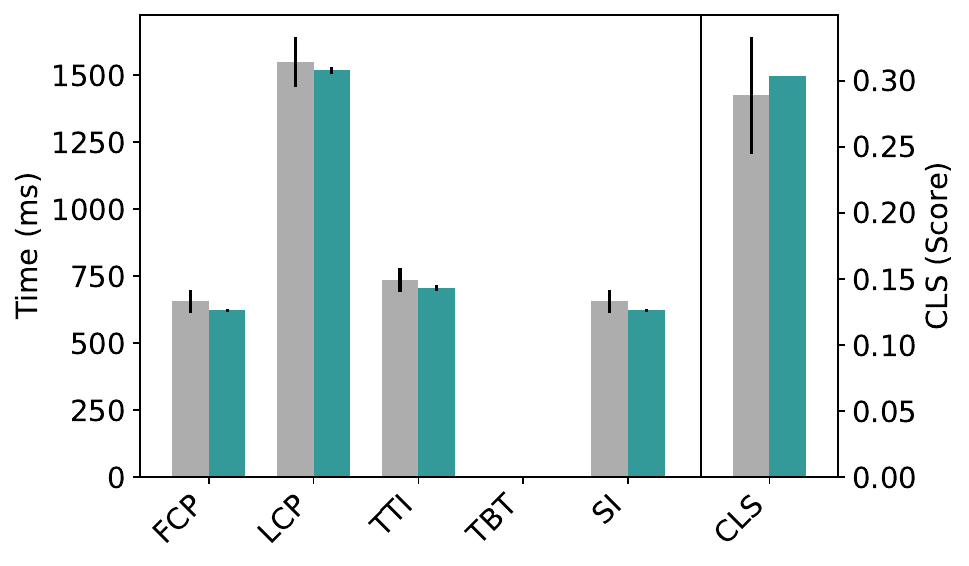}
    \caption{www.apache.com}
    \label{apache}
  \end{subfigure}
  \newline

  \begin{subfigure}[b]{0.44\textwidth}
    \includegraphics[width=\textwidth, height=\figheight, keepaspectratio]{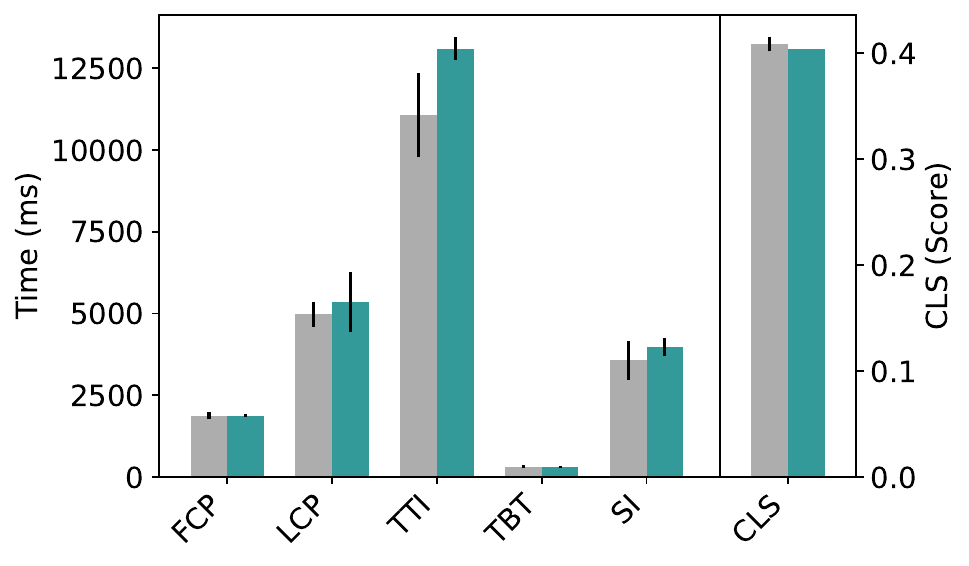}
    \caption{www.nytimes.com}
    \label{nytimes}
  \end{subfigure}
  \hspace{0.02\textwidth} 
  \begin{subfigure}[b]{0.44\textwidth}
    \includegraphics[width=\textwidth, height=\figheight, keepaspectratio]{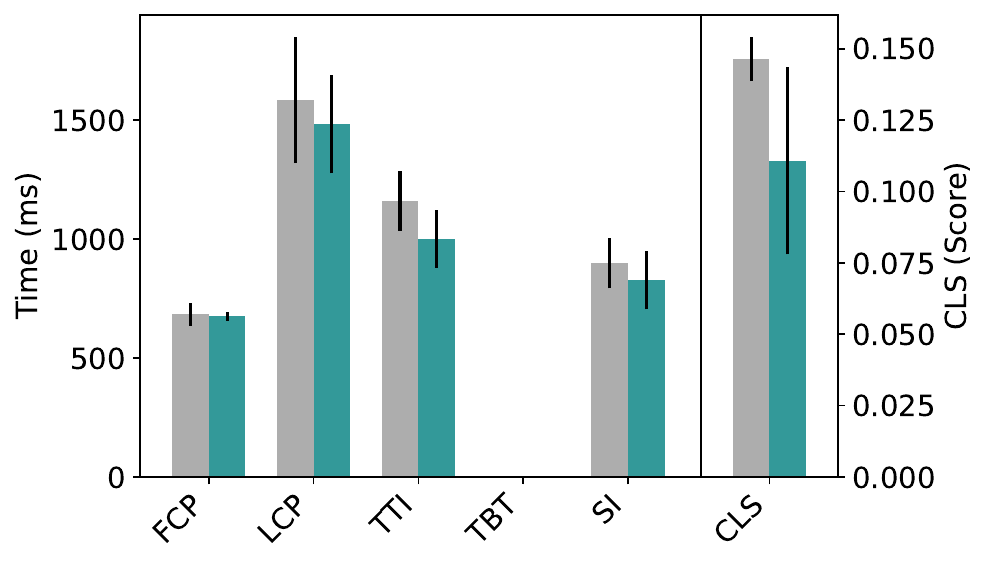}
    \caption{www.statcounter.com}
    \label{statcounter}
  \end{subfigure}
  \newline

  \begin{subfigure}[b]{0.44\textwidth}
    \includegraphics[width=\textwidth, height=\figheight, keepaspectratio]{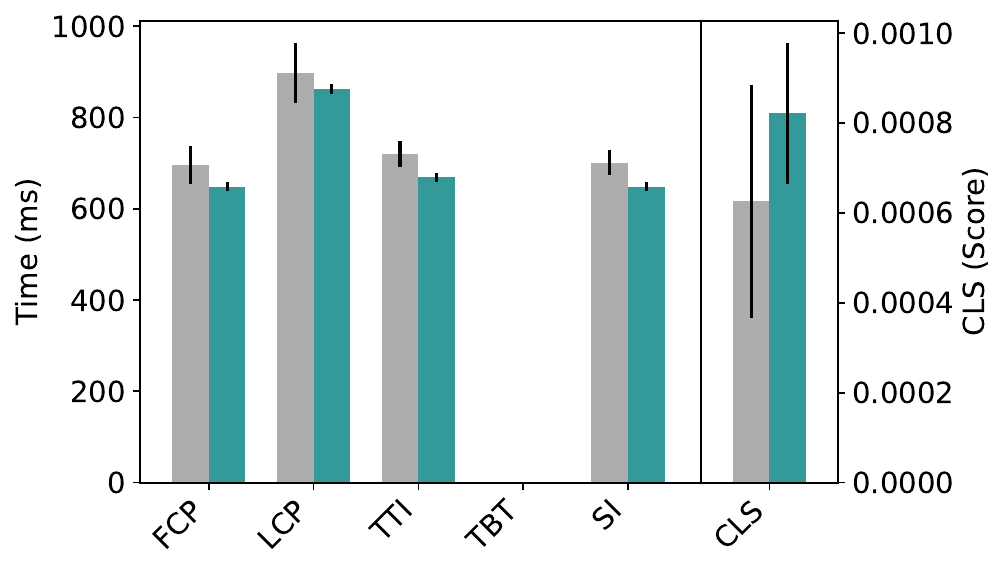}
    \caption{www.w3.org}
    \label{w3c}
  \end{subfigure}
  \hspace{0.02\textwidth} 
  \begin{subfigure}[b]{0.45\textwidth}
    \includegraphics[width=\textwidth, height=\figheight, keepaspectratio]{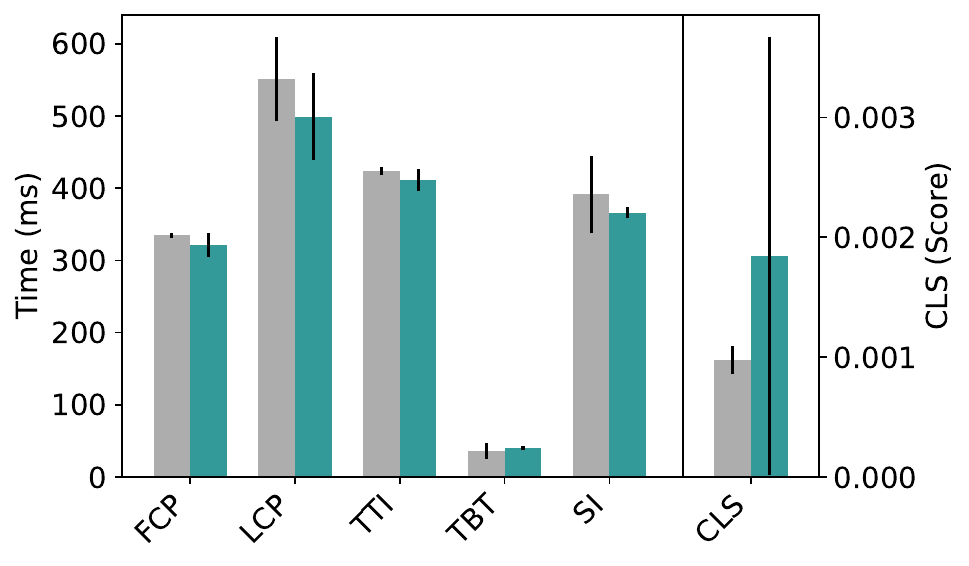}
    \caption{www.wikipedia.org}
    \label{wikipedia}
  \end{subfigure}
  \newline

  \begin{subfigure}[b]{0.44\textwidth}
    \includegraphics[width=\textwidth, height=\figheight, keepaspectratio]{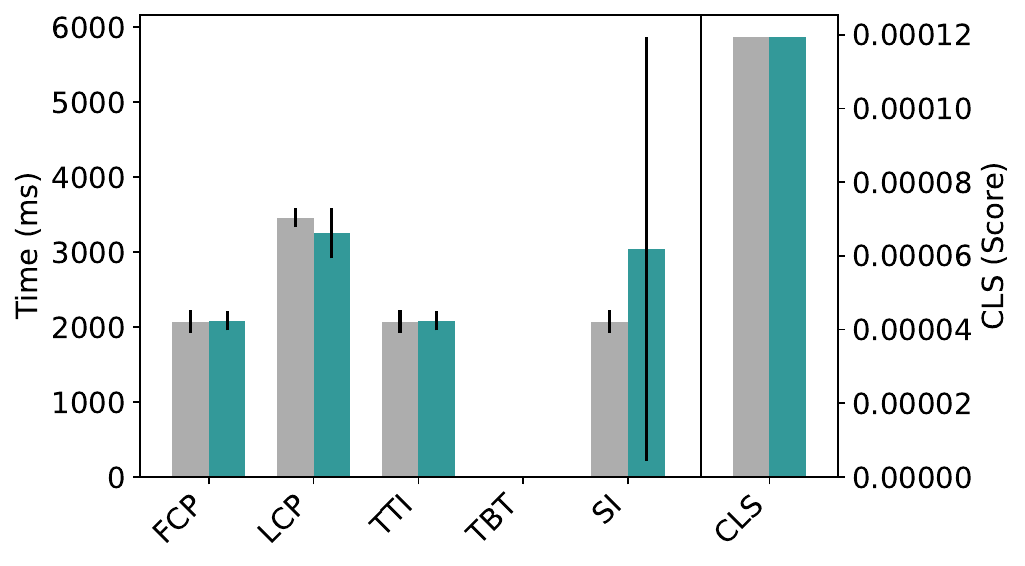}
    \caption{www.apple.com}
    \label{apple}
  \end{subfigure}
  \hspace{0.02\textwidth} 
  \begin{subfigure}[b]{0.44\textwidth}
    \includegraphics[width=\textwidth, height=\figheight, keepaspectratio]{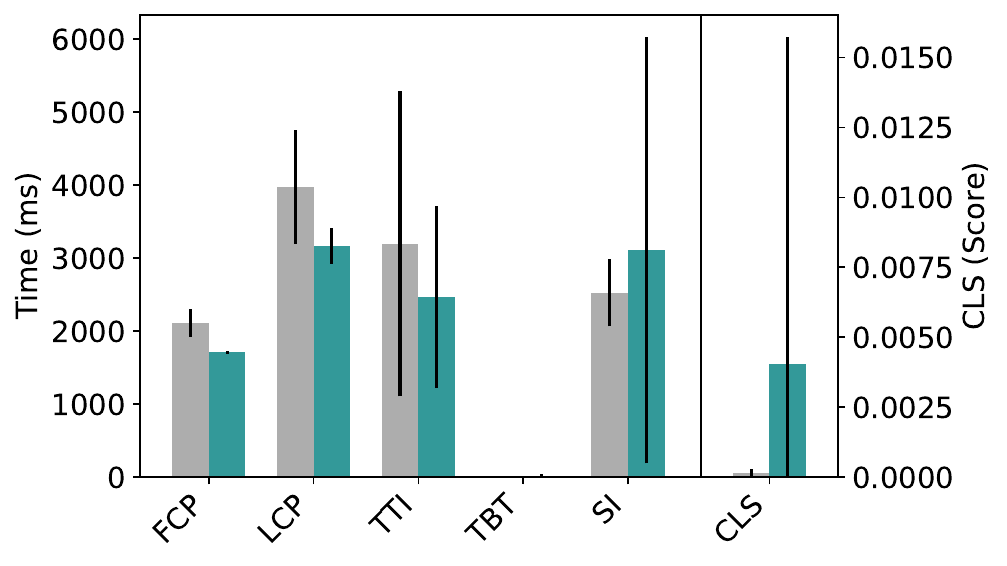}
    \caption{www.etsy.com}
    \label{etsy}
  \end{subfigure}
  \newline

  \includegraphics[width=0.5\textwidth]{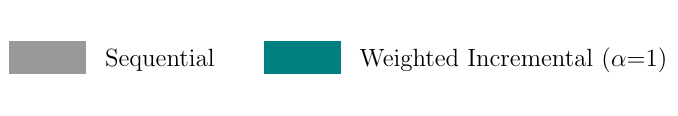} 

  \caption{Comparison of web performance metrics across different websites.}
  \label{barchart}
\end{figure*}

\section{Experimental Evaluation}
\label{sec:results}

Figure~\ref{heatmap} shows the performance difference 
between the weighted incremental mechanism and sequential delivery across various values of $\alpha$. Typically, the most significant performance improvement is observed at $\alpha=1$, the strictest prioritization level, effectively speeding up the loading of content directly visible on the screen. Figure~\ref{barchart} shows the performance of individual websites at $\alpha=1$, featuring dual measurement scales for time-based metrics and CLS, separated by a line and error bars signifying the standard deviation.


However, for some combination of metrics and $\alpha$ values, we observe drop in performance. It should be noted that this is caused by websites that either heavily rely on scripts or feature intricate designs and is not an indication of performance drop in general. 



\subsection{First Contentful Paint and Largest Contentful Paint}

Figure~\ref{heatmap} shows an increase in FCP and LCP performance with $\alpha$, values ranging from 0 to 1. 
Figure~\ref{barchart}, shows that FCP and LCP fared well in most websites. However, the New York Times website stood out as an
exception for LCP. This site's distinct challenge for LCP arises from its use of high-resolution imagery and complex interplay between JavaScript and CSS.


LCP constitutes 25\% of the Lighthouse performance Score, marking it as one of the most crucial metrics for improvement~\cite{lighthouse}. To further illustrate the impact on LCP improvement, our
regression plots in Figure~\ref{regression} show that weighted incremental
mechanism at $\alpha = 1$, leads to shorter LCP load times than
sequential scheduling. This observation aligns with improvements in LCP through prioritization as noted by Cloudflare~\cite{cloudflare2023http3}.
Similar improvements were observed for other metrics but
were omitted here due to space constraints.

\subsection{Time To Interactive and Total Blocking Time}

TTI and TBT remained positive with the increase in alpha values, as shown in Figure~\ref{heatmap}. However, at $\alpha=1$, our Weighted incremental mechanism delayed the downloading of scripts critical for page rendering, which deteriorated TTI performance. In the case of TBT, Etsy's performance at $\alpha$ values of 0.75 and 1 was an exception. The strategy of prioritizing above-the-fold content to enhance FCP and LCP metrics through our weighted incremental mechanism led to a significant increase in TBT for Etsy.

Figure~\ref{barchart} shows that most websites demonstrated favorable TTI scores. However, the TTI score for New York Times was higher, indicating poorer performance, an outcome that can be attributed to New York Time's complex webpage architecture. The TBT scores for Amazon and the New York Times were similar, as illustrated in Figure~\ref{barchart}, while Wikipedia's TBT score experienced a modest increase, signifying a decrease in TBT performance. It should be noted that if TBT is not visible in the Figure~\ref{barchart}, this indicates that its value is either negligible or zero, implying there is no blocking time.

\subsection{Speed Index}

Figure~\ref{heatmap} indicates an improvement in performance for SI at $\alpha$ levels 0, 0.25, and 0.75. However, at  $\alpha = 0.5$ and $\alpha = 1$, a notable decline is observed,  influenced by the outlier performance from Apple, which exhibited a 41\% reduction in SI at both levels. This substantial dip is attributed to the complexity of Apple's website.

The SI fared well under the weighted incremental mechanism over the purely sequential approach, as shown in Figure~\ref{barchart}. Exceptions to this were observed, with the New York Times experiencing a marginal dip. At the same time, Etsy and Apple encountered a more marked decrease in their SI values when employing the weighted incremental mechanism.

This reduction in SI for certain websites aligns with our targeted optimization of the FCP and LCP, since the proposed mechanism prioritizes the loading of content that is visible on the screen, commonly referred to as above-the-fold.


\subsection{Cumulative Layout Shift}

Figure~\ref{heatmap} shows that the weighted incremental approach at $\alpha = 0.25$ led to more consistent CLS performance in comparison to higher $\alpha$ values. Moderate weighting at \(\alpha = 0.5\) and $\alpha = 0.75$ resulted in variable CLS performance, while a full weighting at $\alpha = 1$ corresponded with a 22.42\% decrease in CLS.

Figure~\ref{barchart} shows variability in CLS. Websites such as Wikipedia, W3C, and Etsy exhibited particularly poorer CLS performance across different $\alpha$ values. For Wikipedia and Etsy, the weighted incremental mechanism displayed substantial variance. In contrast, sites like Apple and The New York Times showed minimal variance.

\section{Conclusions}
\label{sec:conclusions}

The EPS fundamentally reshapes the web resource delivery by enabling more adaptable prioritization strategies tailored to web content's dynamic needs. This paper explores the transition from sequential to incremental and eventually to a weighted incremental resource delivery method. This progression allowed us to explore how varying levels of resource prioritization impact web performance across several metrics.

Our weighted incremental mechanism, based on EPS, demonstrates a performance improvement over sequential delivery with respect to Lighthouse performance metrics. For the initial visual cues of website loading, such as FCP and LCP, this approach was beneficial. It ensured that the most crucial content, visible above the fold, loaded quickly. TTI consistently increased across various $\alpha$ values, with medium levels yielding the most notable improvements. 

The TBT also showed improvement with increasing values of $\alpha$, indicating reduced time when the main thread was blocked. SI also reflected improvements with the weighted incremental mechanism. CLS performance was positive at lower alpha levels, suggesting that a less weighted incremental approach fosters fewer layout shifts. 


In our future research, utilizing a variety of HTTP/3 server implementations could yield insightful comparative data on the effectiveness of our prioritization strategy. Augmenting our mechanism with dynamic priority updates may further improve HTTP/3 web content delivery and QoE. Moreover, adopting a hybrid delivery strategy combining incremental and non-incremental resource loading could improve performance.

\bibliographystyle{ieeetr}
\bibliography{bibliography.bib}
\end{document}